\documentclass[journal]{IEEEtran}

\usepackage[utf8]{inputenc}
\usepackage[T1]{fontenc}
\usepackage{amsmath,amssymb,amsfonts}
\usepackage{graphicx}
\usepackage{booktabs}
\usepackage{hyperref}
\usepackage{listings}
\usepackage{xcolor}

\lstdefinestyle{javacode}{
  language=Java,
  basicstyle=\ttfamily\footnotesize,
  keywordstyle=\bfseries,
  stringstyle=\ttfamily,
  commentstyle=\ttfamily,
  showstringspaces=false,
  breaklines=true,
  frame=single,
  framerule=0.4pt,
  rulecolor=\color{black},
  backgroundcolor=\color{gray!7},
  columns=fullflexible,
  keepspaces=true,
  tabsize=2,
  xleftmargin=4pt,
  xrightmargin=4pt,
  framexleftmargin=4pt,
  framexrightmargin=4pt,
  aboveskip=3pt,
  belowskip=3pt
}

\usepackage{balance}
\usepackage{multirow}
\usepackage{array}
\usepackage{url}
\usepackage{cite}
\usepackage{tabularx}
\usepackage{algorithm}
\usepackage{algpseudocode}
\usepackage{textcomp}
\usepackage{orcidlink}

\hypersetup{
  colorlinks = true,
  linkcolor  = blue!70!black,
  citecolor  = blue!70!black,
  urlcolor   = blue!60!black
}

\graphicspath{{figures/}{../figures/}}

\begin{document}

\title{QASecClaw: A Multi-Agent LLM Approach for False Positive Reduction in Static Application Security Testing}

\author{%
Mohd~Ruhul~Ameen\IEEEauthorrefmark{2}, Md~Takrim~Ul~Alam\IEEEauthorrefmark{1}, and~Akif~Islam\IEEEauthorrefmark{2}\\
\IEEEauthorrefmark{1}Department of Information and Communication Engineering, University of Rajshahi, Bangladesh\\
\IEEEauthorrefmark{2}Department of Computer Science and Engineering, University of Rajshahi, Bangladesh\\
E-mails: ameen@marshall.edu; takrimulalom@gmail.com; iamakifislam@gmail.com
}


\maketitle



\begin{abstract}
Static Application Security Testing (SAST) tools help developers identify security vulnerabilities before software is released. However, these tools often report many false positives, which increases manual review effort and reduces developer trust. When developers repeatedly encounter incorrect warnings, they may start skipping or ignoring scanner output. As a result, real vulnerabilities can remain hidden among false positive reports and may reach production unnoticed. We present \textbf{QASecClaw}, a multi-agent approach that combines conventional SAST with coding-specialized Large Language Model (LLM)-based contextual code review. In this approach, a SAST engine first reports candidate vulnerabilities, and a coding-specialized LLM-based SAST Filter Agent then reviews each finding with source-code context to determine whether it is likely to be a true positive or a false positive. QASecClaw uses specialized agents, including a Test Planning Agent, a Security Validation Agent backed by Semgrep, an Evidence Correlation Agent, and a SAST Filter Agent, all coordinated by a Mission Orchestrator. We evaluate QASecClaw on the complete OWASP Benchmark v1.2, which contains 2,740 Java test cases across 11 Common Weakness Enumeration (CWE) categories with ground-truth labels. QASecClaw achieves an F1-score of 90.93\%, compared with 78.39\% for standalone Semgrep. This improvement is mainly driven by an 88.6\% reduction in false positives, from 560 to 64, with only a 3.1\% reduction in recall. Per-CWE analysis shows strong gains on injection-related vulnerabilities, including SQL Injection with an F1-score of 94.05\% and Cross-Site Scripting with an F1-score of 89.58\%, while maintaining near-perfect performance on Weak Cryptography with an F1-score of 99.61\%. These results show that LLM-augmented multi-agent verification is a practical approach for making SAST output more accurate, useful, and trustworthy. The source code and experimental scripts are publicly available to support reproducibility.
\end{abstract}

\begin{IEEEkeywords}
Multi-agent systems, large language models, static application security testing, false positive reduction, software security, OWASP benchmark, applied artificial intelligence, secure software engineering, AI agents.
\end{IEEEkeywords}

\section{Introduction}
\label{sec:introduction}

\IEEEPARstart{S}{oftware} now sits at the center of daily life. Banking, healthcare, education, communication, transportation, government services, and business operations all depend on applications that are updated continuously. In the Artificial Intelligence (AI) era, this dependence is increasing further, as software systems are no longer only storing and processing data but also assisting decisions, automating workflows, and interacting with users through intelligent agents. As a result, the security of software code has become a practical necessity rather than an optional engineering concern. A single vulnerability can expose sensitive data, disrupt critical services, or become an entry point for a larger attack. The growing economic cost of cybercrime reflects this reality \cite{morgan2023cost}.

To reduce these risks, organizations commonly use Static Application Security Testing (SAST) tools during development. SAST tools inspect source code without executing the program and report code patterns that may correspond to known security weaknesses \cite{chess2007secure}. This makes them attractive for early security checking because they can be integrated into development pipelines and can warn developers before vulnerable code reaches production. In principle, SAST provides a valuable first line of defense.

The difficulty begins when SAST tools are used in practice. Most SAST engines are intentionally conservative: they prefer to report a suspicious pattern rather than risk missing a real vulnerability. For example, if user input appears to flow into a Structured Query Language (SQL) query, a shell command, a file path, or an output response, the tool may raise an alert. This behavior helps preserve recall, but it also creates many false positives. A finding may look dangerous to a pattern matcher even though the code is safe because the input has already been sanitized, encoded, validated, or passed through a parameterized Application Programming Interface (API). Fig.~\ref{fig:sast_false_positive_example} illustrates this problem with a simplified Java example. The difficulty is that the same syntactic pattern may represent either a real vulnerability or a safe implementation, depending on the surrounding code context.

\begin{figure*}[!t]
\centering

\begin{minipage}[t]{0.47\textwidth}
\centering
\textbf{Case A: Likely true positive}

\vspace{1mm}

\begin{lstlisting}[style=javacode]
String userId =
    request.getParameter("id");

String query =
    "SELECT * FROM users WHERE id = "
    + userId;

stmt.execute(query);
\end{lstlisting}

\vspace{1mm}

\footnotesize
User input flows directly into a SQL query through string concatenation. A SAST warning is likely valid because no sanitization or parameterization is visible.
\end{minipage}
\hfill
\begin{minipage}[t]{0.47\textwidth}
\centering
\textbf{Case B: Likely false positive}

\vspace{1mm}

\begin{lstlisting}[style=javacode]
String userId =
    request.getParameter("id");

userId =
    sanitizeAsInteger(userId);

String query =
    "SELECT * FROM users WHERE id = "
    + userId;

stmt.execute(query);
\end{lstlisting}

\vspace{1mm}

\footnotesize
The same syntactic pattern appears, but the input is sanitized before reaching the SQL sink. Contextual review can suppress this warning as a likely false positive.
\end{minipage}

\vspace{2mm}

\caption{Simplified Java example showing why SAST false positives occur. A pattern-based scanner may flag both snippets because user input appears near a SQL sink. The key difference is contextual: in Case B, the input is sanitized before use. QASecClaw uses a coding-specialized LLM-based SAST Filter Agent to review such context before retaining or suppressing a finding.}
\label{fig:sast_false_positive_example}
\end{figure*}

This false positive problem is not only a technical limitation, it is also a human problem. Every warning requires developer or security analyst attention. When a scan produces hundreds of alerts and many of them are incorrect, developers gradually lose trust in the tool. They may skip warnings, disable noisy rules, or ignore the scanner output entirely. This is dangerous because real vulnerabilities can remain hidden among false positive reports and pass into production unnoticed. Prior work has shown that developers often avoid static analysis tools when warnings are too noisy or difficult to act on \cite{johnson2013don}. Industrial experience similarly shows that static analysis is useful only when its output is relevant, actionable, and trusted by developers \cite{sadowski2015tricorder}.

Existing solutions only partially address this issue. More precise program analysis can improve source to sink reasoning, but it can still struggle with custom sanitizers, framework behavior, callbacks, and application specific logic. Dynamic validation can provide stronger evidence, but it requires a runnable system, meaningful inputs, and sufficient test coverage. Machine learning based alert prioritization can reduce review effort, but it often depends on labeled project history and may not transfer well across repositories or vulnerability categories. These limitations motivate a different direction: instead of replacing SAST, we can keep its strength as a high recall scanner and add a contextual reviewer that checks whether each reported finding is truly exploitable.

Large Language Models (LLMs) provide a promising way to perform this contextual review. Modern coding specialized LLMs can read source code, follow data movement, recognize common sanitization patterns, and reason about whether a reported source to sink path is actually dangerous \cite{chen2021evaluating}. However, using an LLM as the primary vulnerability detector is risky because the model may hallucinate vulnerabilities or produce inconsistent judgments. The safer design is to use the LLM in a narrower role: not to search the entire codebase from scratch, but to review candidate findings already produced by a SAST engine.

In this paper, we present \textbf{QASecClaw}, a multi agent approach for reducing false positives in SAST output. QASecClaw combines conventional SAST scanning with coding specialized LLM based contextual code review. A SAST engine first produces candidate findings. Then, a SAST Filter Agent reviews each finding with source code context and decides whether the finding is likely to be a true positive or a false positive. The broader approach is organized through specialized agents, including a Test Planning Agent, a Security Validation Agent, an Evidence Correlation Agent, and a SAST Filter Agent, coordinated by a central Mission Orchestrator. This decomposition follows the multi agent systems principle of assigning complex tasks to specialized cooperating components \cite{wooldridge2009introduction}.

The core idea is simple: SAST should continue to cast a wide net, while the LLM should help decide which reported findings deserve developer attention. This division of labor matches how human security triage works. Security analysts do not normally search an entire codebase from scratch; they inspect candidate findings, examine the surrounding code, and decide whether each warning is real. QASecClaw automates this review step while using a conservative fail open policy: if the LLM fails, times out, or returns malformed output, the original SAST finding is retained rather than suppressed.

The contributions of this paper are as follows.

\textbf{1)} We propose \textbf{QASecClaw}, a multi agent approach that combines conventional SAST scanning with coding specialized LLM based contextual review to reduce false positives in security analysis.

\textbf{2)} We introduce a \textbf{SAST Filter Agent} that reviews SAST findings using source code context and Common Weakness Enumeration (CWE) specific reasoning to classify findings as true positives or false positives.

\textbf{3)} We incorporate a conservative \textbf{fail open mechanism} so that LLM failures, timeouts, or malformed responses do not silently suppress potentially real vulnerabilities.

\textbf{4)} We evaluate the approach on the complete Open Worldwide Application Security Project (OWASP) Benchmark v1.2 and compare it with standalone Semgrep to assess false positive reduction and detection quality.

\noindent\textbf{Code Availability:}
The implementation of QASecClaw and supporting experimental scripts are publicly available at:
\url{https://github.com/takrim1999/qasecqlaw}.

\section{Related Work}
\label{sec:related}

\subsection{Static Application Security Testing}
\label{sec:rw_sast}

Static Application Security Testing (SAST) tools analyze source code without executing the program. Tools such as Semgrep, SonarQube, SpotBugs, Fortify, and Coverity are commonly used to detect potential vulnerabilities during development. These tools typically rely on rule matching, Abstract Syntax Tree (AST) analysis, taint tracking, and data-flow rules to identify suspicious code patterns \cite{semgrep2023}. SonarQube provides a broader code-quality and security-analysis platform \cite{sonarqube2023}, while SpotBugs focuses mainly on Java bytecode-level bug patterns \cite{spotbugs2023}. Earlier static-analysis work also established SAST as an important practice for secure programming \cite{chess2007secure}.

The main strength of SAST is early detection. A scanner can flag possible vulnerabilities before deployment, which makes it useful for Continuous Integration (CI) pipelines. However, most SAST tools are intentionally conservative: they prefer to report a suspicious pattern rather than miss a real vulnerability. This improves recall but creates many false positives. For example, a tool may report SQL Injection when user input appears near a database query, even if the input has already been sanitized or the query uses a parameterized Application Programming Interface (API). This gap between syntactic warning and semantic exploitability is the main limitation of conventional SAST.

This limitation also affects developer behavior. Johnson et al. showed that developers often avoid static-analysis tools when warnings are noisy, hard to interpret, or poorly integrated into their workflow \cite{johnson2013don}. Sadowski et al. similarly showed that static analysis at industrial scale depends not only on finding defects, but also on presenting results that developers can trust and act on \cite{sadowski2015tricorder}. Therefore, improving SAST requires more than increasing the number of reported findings; it requires making those findings more precise and useful.

\subsection{False Positive Reduction Techniques}
\label{sec:rw_fp}

A major direction for reducing SAST false positives is stronger program analysis. Frameworks such as Soot support deeper Java analysis, including control-flow and data-flow reasoning across methods \cite{lam2011soot}. Such techniques improve source-to-sink tracking, but they still struggle with reflection, callbacks, framework-specific behavior, custom sanitizers, and complex library interactions. In practice, increasing static-analysis precision can also increase computational cost and still may not capture the application-specific context needed to distinguish safe code from vulnerable code.

Hybrid static and dynamic approaches attempt to validate static findings using runtime behavior. The intuition is simple: if a static warning is supported by runtime evidence, it is more likely to be real. This can improve confidence, but it assumes that the application can be executed, configured, and tested with meaningful inputs. Beyer emphasized the importance of validating software-analysis results, but such validation often requires conditions that are not available in early-stage SAST scanning \cite{beyer2017software}. As a result, hybrid methods are useful when runtime evidence exists, but they are less suitable for purely static benchmark settings or early CI scans.

Machine-learning-based alert classification has also been used to prioritize or suppress warnings. Heckman and Williams reviewed techniques for identifying actionable static-analysis alerts and showed that learning-based methods can reduce manual review effort \cite{heckman2011systematic}. Flynn et al. trained classifiers to prioritize alerts from multiple static-analysis tools \cite{flynn2018prioritizing}. Ko\c{c} et al. also studied classifiers for false positive reports emitted by static-analysis tools \cite{koc2017learning}. These approaches are useful when sufficient historical data is available, but they often depend on project-specific features and labeled examples. When the codebase, development team, tool configuration, or vulnerability class changes, the learned model may not generalize well.

Deep-learning approaches move beyond handcrafted features by learning vulnerability patterns directly from code. VulDeePecker introduced neural vulnerability detection using code gadgets \cite{li2018vuldeepecker}. $\mu$VulDeePecker extended this idea to multiclass vulnerability detection \cite{zou2019mu}. Devign used graph neural networks to model program semantics \cite{zhou2019devign}, and LineVul applied transformer models to line-level vulnerability prediction \cite{fu2022linevul}. These models show that neural networks can capture patterns that are difficult to encode manually. However, they require large labeled datasets and can perform poorly when evaluated outside their training distribution. Chakraborty et al. showed that deep-learning vulnerability detectors often struggle to generalize to realistic settings \cite{chakraborty2022deep}. Steenhoek et al. similarly reported that performance varies substantially across datasets and evaluation setups \cite{steenhoek2023empirical}.

Overall, prior false-positive reduction techniques have improved SAST in important ways, but they remain limited. Stronger static analysis can be incomplete or expensive, hybrid validation needs executable systems, machine-learning classifiers need labeled historical data, and deep-learning detectors require large representative datasets. These limitations motivate a method that can reason over code context without requiring task-specific training for every project.

\subsection{Large Language Models for Code Security}
\label{sec:rw_llm}

Large Language Models (LLMs) have introduced new possibilities for code understanding. Earlier models such as CodeBERT and CodeT5 showed that pre-training on source code and natural language can support code search, summarization, and generation \cite{feng2020codebert}. Code Llama and Qwen further extended code-oriented reasoning capabilities \cite{roziere2023code}. Qwen2 improved general and code-related performance across a broader set of tasks \cite{yang2024qwen2}. These models are relevant to security analysis because many vulnerabilities require understanding both program structure and natural-language intent.

LLMs have also been applied to software security tasks. Pearce et al. examined zero-shot vulnerability repair and showed that LLMs can sometimes generate useful fixes, but their outputs require careful checking \cite{pearce2023examining}. Xia et al. studied automated program repair using large pre-trained models \cite{xia2023keep}. Li et al. discussed how LLMs can support program analysis by combining code reasoning with natural-language explanations \cite{li2023hitchhiker}. These studies suggest that LLMs can assist security workflows, especially when the task requires contextual interpretation.

However, direct LLM-based vulnerability detection remains unreliable. When asked to inspect code from scratch, an LLM may hallucinate vulnerabilities, miss subtle flows, or produce inconsistent judgments. Khare et al. showed that LLM vulnerability-detection performance depends strongly on prompt design, context, and vulnerability type \cite{khare2023understanding}. Cheshkov et al. also found that ChatGPT can detect some vulnerabilities but is not reliable enough to replace dedicated security-analysis tools \cite{cheshkov2023evaluation}. LLM4Vuln further highlights that LLM-based vulnerability reasoning requires careful decomposition and evaluation \cite{sun2024llm4vuln}.

QASecClaw follows a different design. It does not use the LLM as a primary detector. Instead, it uses a coding-specialized LLM as a secondary verifier for findings already generated by a high-recall SAST engine. This makes the task narrower and safer. The LLM does not need to search the entire codebase for unknown vulnerabilities; it only needs to review a reported finding, inspect the relevant source-code context, and decide whether the finding is likely to be a true positive or a false positive.

\subsection{Multi-Agent Systems for Software Engineering}
\label{sec:rw_mas}

Multi-agent systems decompose complex tasks into specialized agents that coordinate through shared goals and communication protocols \cite{wooldridge2009introduction}. This idea has recently become prominent in LLM-based software engineering. MetaGPT assigns structured software-development roles to agents, such as product manager, architect, and engineer \cite{hong2024metagpt}. ChatDev models software construction as communication among role-specific agents \cite{qian2024chatdev}. AutoGen provides a general framework for multi-agent conversation and tool use \cite{wu2023autogen}.

Most existing LLM-agent systems focus on generative tasks such as writing code, producing documents, or simulating collaboration. Security verification is different. A generated artifact can be revised later, but an incorrectly suppressed vulnerability can create real risk. Therefore, security-oriented multi-agent systems must be evidence-driven, conservative, and traceable. QASecClaw applies multi-agent orchestration to this verification setting by separating the workflow into planning, security validation, evidence correlation, LLM-based filtering, and reporting.

\subsection{Research Gap}
\label{sec:rw_gap}

Existing work leaves a clear gap. Traditional SAST tools provide broad coverage, but they produce many false positives because they often lack semantic context. Program-analysis and hybrid techniques improve precision, but they can be expensive or require runtime evidence. Machine-learning and deep-learning approaches can prioritize or detect vulnerabilities, but they depend on labeled data and may not generalize well. Direct LLM-based vulnerability detection is promising, but it suffers from hallucination, inconsistency, and uncertain recall.

The underexplored middle path is to keep conventional SAST as the high-recall candidate generator and use an LLM only as a constrained verifier for the findings already produced by SAST. This changes the task from open-ended vulnerability discovery to focused false-positive reduction. QASecClaw addresses this gap by combining Semgrep-based SAST, a coding-specialized LLM-based SAST Filter Agent, evidence correlation, and multi-agent orchestration. The goal is not to replace SAST, but to make SAST output more accurate, actionable, and trustworthy.

\section{QASecClaw Approach}
\label{sec:architecture}

\begin{figure}[!t]
  \centering
  \includegraphics[width=\columnwidth]{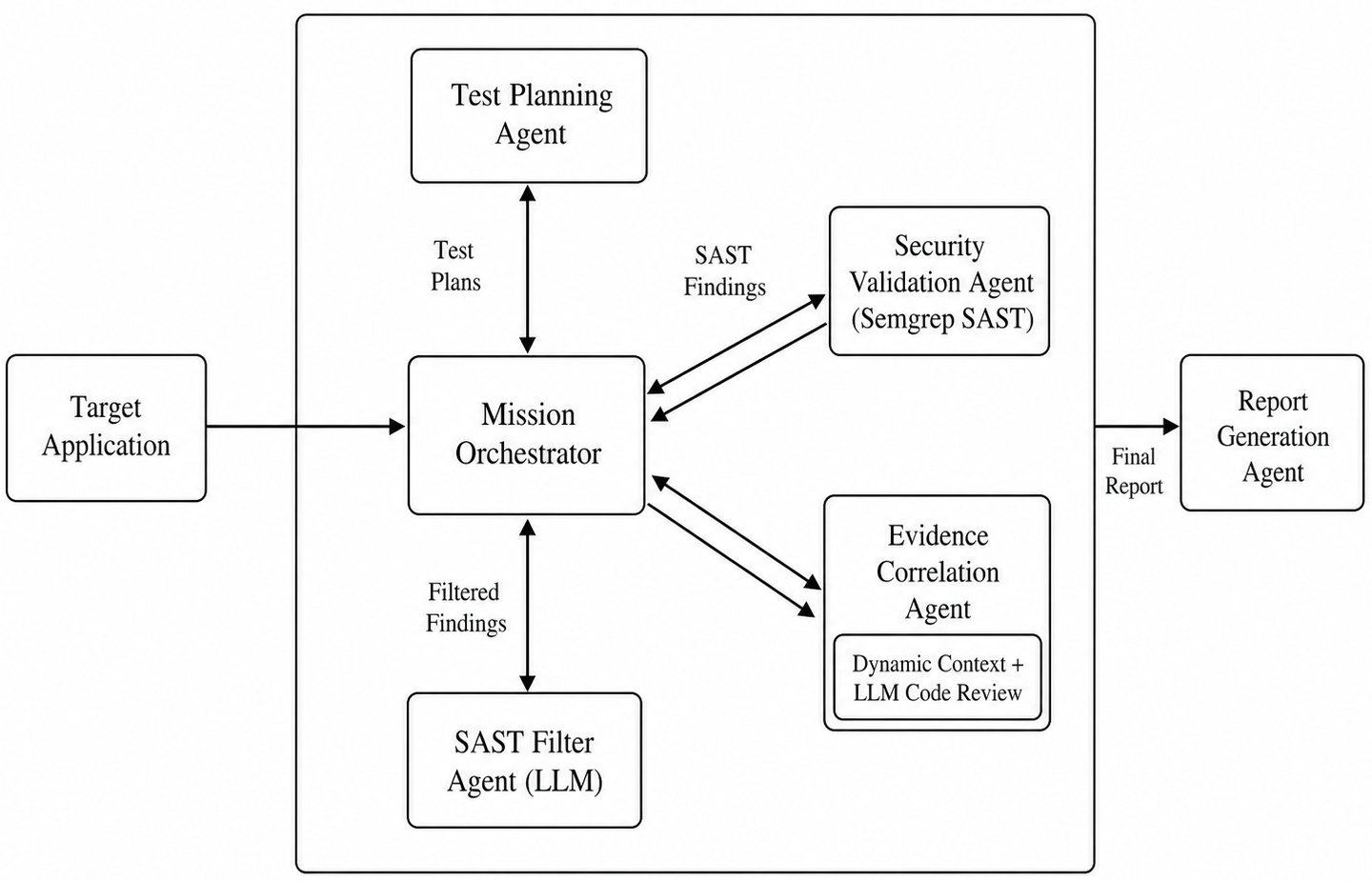}
  \caption{Overview of the QASecClaw multi-agent approach. The Mission Orchestrator coordinates the Test Planning Agent, Security Validation Agent, Evidence Correlation Agent, SAST Filter Agent, and Report Generation Agent.}
  \label{fig:architecture}
\end{figure}

QASecClaw is designed as a multi-agent approach for reducing false positives in Static Application Security Testing (SAST) output. The approach is organized around a central Mission Orchestrator and a set of specialized agents. Each agent performs a specific role in the security assessment pipeline, while the Mission Orchestrator manages the overall analysis flow. Fig.~\ref{fig:architecture} shows the overall architecture.

\subsection{Mission Orchestrator}

The Mission Orchestrator controls the analysis lifecycle. It receives the target application, initializes the run, coordinates the agents, and passes outputs from one stage to the next. In the current implementation, the orchestrator first supports test planning, then invokes the Security Validation Agent to run SAST analysis, sends findings to the Evidence Correlation Agent, and finally produces the verified report through the Report Generation Agent.

\subsection{Security Validation Agent}

The Security Validation Agent is responsible for conventional SAST scanning. In this study, it uses Semgrep as the underlying scanner. Semgrep produces candidate security findings from the target source code, and these findings are normalized into a common format containing the vulnerability type, file path, line information, severity, and description. This normalization allows later agents to process findings independently of the specific scanner.

\subsection{Evidence Correlation Agent}

The Evidence Correlation Agent receives SAST findings and determines whether additional evidence is available. In live applications, this evidence may include User Interface (UI) traces, Application Programming Interface (API) logs, runtime behavior, or system anomalies. Findings supported by dynamic evidence can be treated as verified. Findings without such evidence are forwarded to the SAST Filter Agent for contextual code review. Since the OWASP Benchmark is a static benchmark, the dynamic evidence path is not active in this evaluation; therefore, the reported results mainly evaluate the SAST filtering pathway.

\subsection{SAST Filter Agent}
\label{sec:sast_filter_agent}

\begin{figure}[!t]
  \centering
  \includegraphics[width=0.78\columnwidth]{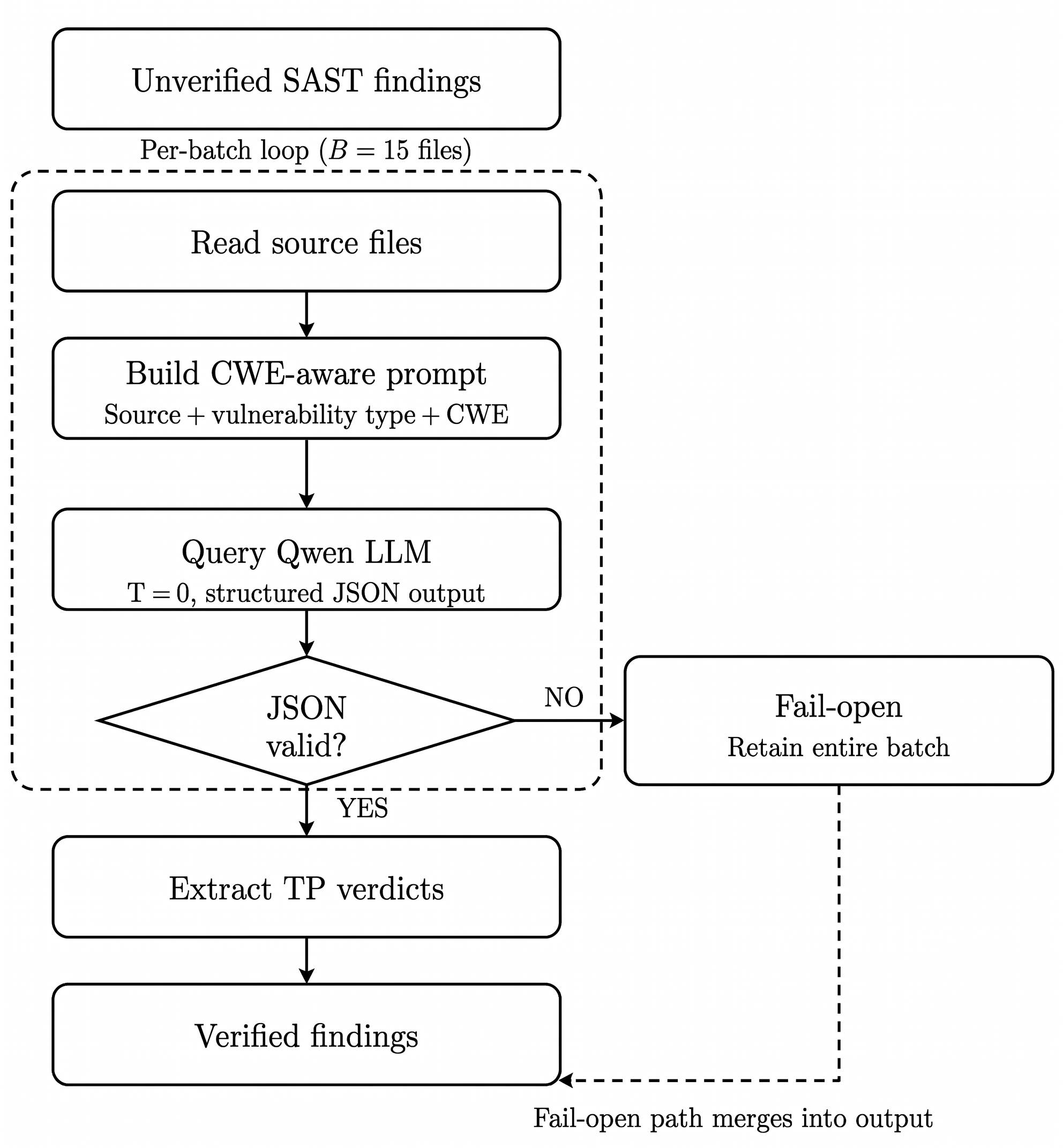}
  \caption{Workflow of the SAST Filter Agent. Unverified SAST findings are processed in batches of 15 files. For each batch, the agent reads source files, builds a CWE-aware prompt, queries the coding-specialized LLM, validates the structured JSON response, and extracts true-positive verdicts. If the response is invalid, the fail-open path retains the entire batch.}
  \label{fig:filter_workflow}
\end{figure}

The SAST Filter Agent is the main component responsible for false positive reduction. It reviews unverified SAST findings using source-code context and a coding-specialized Large Language Model (LLM). For each finding, the agent reads the relevant source file, builds a prompt containing the vulnerability type, Common Weakness Enumeration (CWE) category, file location, and source code, and asks the LLM to determine whether the finding is likely to be a true positive or a false positive. Fig.~\ref{fig:filter_workflow} illustrates the internal workflow of the SAST Filter Agent, including batching, prompt construction, LLM querying, JSON validation, and fail-open handling.

The filtering process is intentionally conservative. If the LLM returns a valid structured response, only findings classified as true positives are retained. If the LLM call fails, times out, or returns malformed JSON, QASecClaw applies a fail-open policy and retains the entire batch. This prevents LLM failure from silently suppressing real vulnerabilities.

\subsection{Report Generation Agent}

The Report Generation Agent produces the final output of the analysis. It receives the verified findings from the Evidence Correlation Agent and prepares a concise report containing retained vulnerabilities and suppressed false positives. This makes the final SAST output easier for developers and security analysts to review.

\section{Experimental Methodology}
\label{sec:setup}

\subsection{Research Questions}
We investigate the following research questions:

\begin{enumerate}
    \item \textbf{RQ1}: How does QASecClaw compare with standalone Semgrep on the OWASP Benchmark in terms of precision, recall, F\textsubscript{1}-score, false positive rate, and Youden's $J$?
    \item \textbf{RQ2}: To what extent does QASecClaw's LLM-augmented filtering reduce false positives compared with standalone Semgrep?
    \item \textbf{RQ3}: How does QASecClaw's detection performance vary across Common Weakness Enumeration (CWE) categories?
\end{enumerate}

\subsection{Dataset: OWASP Benchmark v1.2}
\label{sec:dataset}

We use the \textbf{OWASP Benchmark Project v1.2} as the evaluation dataset \cite{owasp2023benchmark}. It is a free and open benchmark designed to measure how accurately automated security tools detect software vulnerabilities. The benchmark is suitable for this study because it contains both vulnerable programs and safe programs that look similar to vulnerable ones, which allows us to evaluate not only detection performance but also false positive behavior.

OWASP Benchmark v1.2 contains \textbf{2,740 Java test cases}. Each test case is implemented as a self-contained servlet and is associated with a specific Common Weakness Enumeration (CWE) category. Some test cases contain real exploitable vulnerabilities, while others are intentionally written to resemble vulnerable patterns without actually being exploitable. These safe cases are important because they test whether a tool can distinguish a real vulnerability from code that only appears suspicious.

Each test case is identified by a filename in the format \texttt{BenchmarkTestNNNNN.java}. The official ground truth is provided in \texttt{expectedresults-1.2.csv}, which maps each test case to its CWE category and a Boolean label indicating whether the case is truly vulnerable. Table~\ref{tab:dataset} shows the distribution of vulnerable and safe test cases across the CWE categories in OWASP Benchmark v1.2.

\begin{table}[t]
\centering
\caption{OWASP Benchmark v1.2 Dataset Distribution}
\label{tab:dataset}
\begin{tabular}{llrrr}
\toprule
\textbf{CWE} & \textbf{Category} & \textbf{Vuln.} & \textbf{Safe} & \textbf{Total} \\
\midrule
CWE-22  & Path Traversal           & 133  & 135  & 268 \\
CWE-78  & Command Injection        & 126  & 125  & 251 \\
CWE-79  & Cross-Site Scripting     & 246  & 209  & 455 \\
CWE-89  & SQL Injection            & 272  & 232  & 504 \\
CWE-90  & LDAP Injection           & 27   & 32   & 59  \\
CWE-327 & Weak Cryptography        & 130  & 116  & 246 \\
CWE-328 & Weak Hashing             & 129  & 107  & 236 \\
CWE-330 & Weak Randomness          & 218  & 275  & 493 \\
CWE-501 & Trust Boundary Violation & 83   & 43   & 126 \\
CWE-614 & Insecure Cookie          & 36   & 31   & 67  \\
CWE-643 & XPath Injection          & 15   & 20   & 35  \\
\midrule
\multicolumn{2}{l}{\textbf{Total}} & \textbf{1,415} & \textbf{1,325} & \textbf{2,740} \\
\bottomrule
\end{tabular}
\end{table}

\subsection{Baseline}
We use \textbf{Semgrep Community Edition}~\cite{semgrep2023} as the SAST baseline, configured with the \texttt{auto} ruleset (which includes the \texttt{p/default} and community rulesets). Semgrep was selected because: (1) it is a widely-adopted open-source SAST tool, (2) it supports Java analysis, and (3) it produces structured JSON output amenable to automated evaluation.

\subsection{Evaluation Metrics}
We adopt standard information retrieval metrics used in the security tool evaluation literature~\cite{delaitre2015evaluating}:
\begin{align}
    \text{Precision} &= \frac{TP}{TP + FP} \label{eq:precision} \\
    \text{Recall (TPR)} &= \frac{TP}{TP + FN} \label{eq:recall} \\
    \text{F1-Score} &= 2 \times \frac{\text{Precision} \times \text{Recall}}{\text{Precision} + \text{Recall}} \label{eq:f1} \\
    \text{FPR} &= \frac{FP}{FP + TN} \label{eq:fpr} \\
    \text{Youden's \textit{J}} &= \text{TPR} - \text{FPR} \label{eq:youden}
\end{align}

Youden's J statistic~\cite{youden1950index} is the OWASP Benchmark Project's primary ranking metric, as it captures both detection and false alarm rates in a single value.

\subsection{Experimental Procedure}
The evaluation was conducted in a two-stage pipeline:
\begin{enumerate}
    \item \textbf{Baseline assessment}: Standalone Semgrep was executed against the full OWASP Benchmark test code directory. The JSON output was parsed and normalized to extract unique test case findings.
    \item \textbf{QASecClaw assessment}: The complete QASecClaw pipeline was executed, including Semgrep scanning followed by the SAST Filter Agent's LLM-based code review for all uncorrelated findings.
\end{enumerate}

Both tools' findings were cross-referenced against the OWASP ground truth at the file level (\texttt{BenchmarkTestNNNNN}) to compute confusion matrix entries and derived metrics.

\subsection{Implementation Details}
QASecClaw is implemented in TypeScript running on Node.js, and the source code with supporting experimental scripts is publicly available at \url{https://github.com/takrim1999/qasecqlaw}. The SAST Filter Agent uses the \textbf{Qwen 3.5 Plus} model via API for code review. The SAST Filter Agent uses the \textbf{Qwen 3.5 Plus} model via API for code review. Findings are processed in batches of 15 files to manage API throughput and context limitations. The complete pipeline execution (including Semgrep scanning, LLM-based filtering, and scoring) completed in approximately \textbf{45 minutes} on a standard development workstation.

\section{Results}
\label{sec:results}

\subsection{RQ1: Overall Performance Comparison}

\begin{figure}[!t]
  \centering
  \includegraphics[width=\columnwidth]{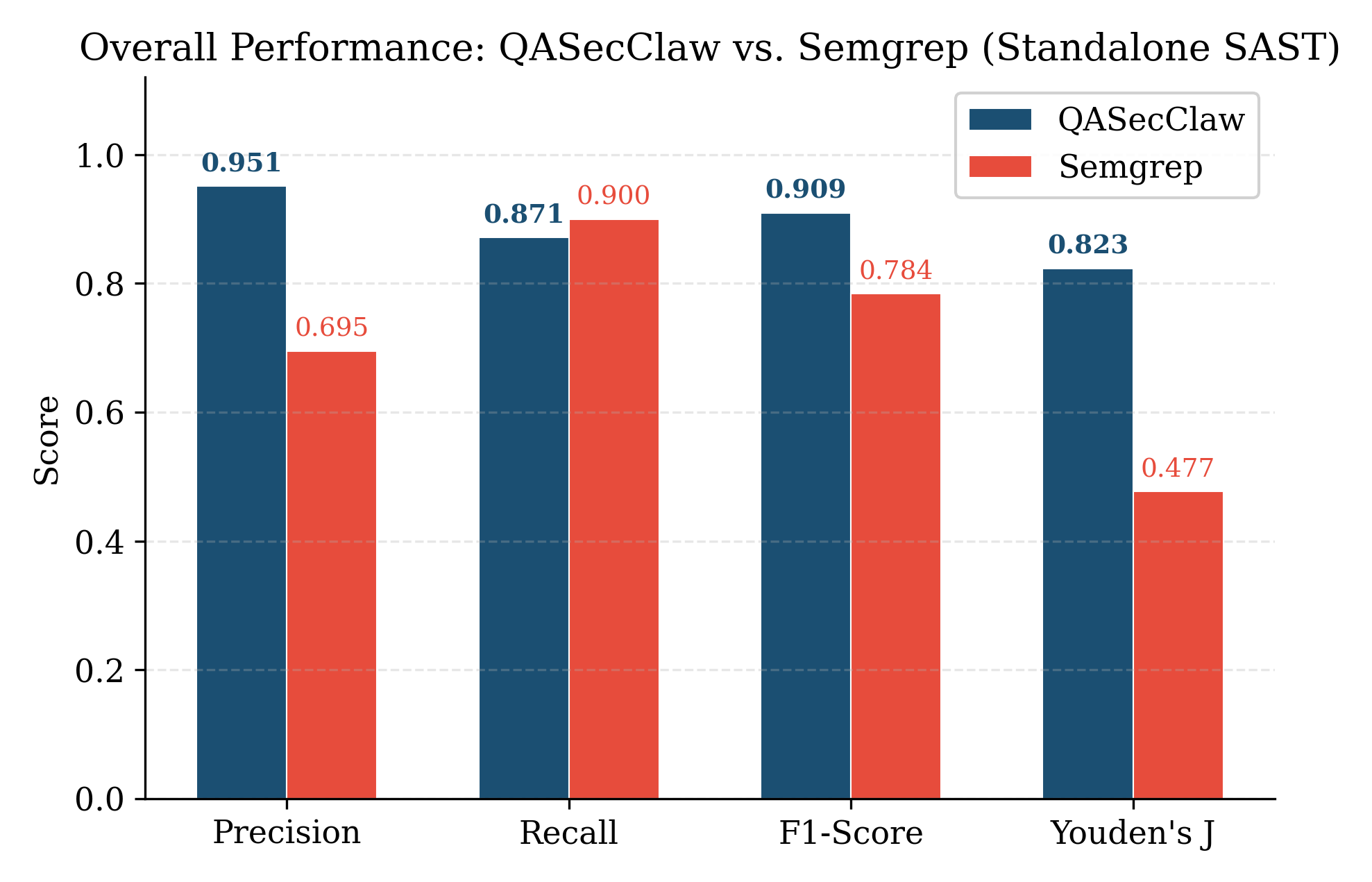}
  \caption{Aggregate metric comparison between QASecClaw and standalone Semgrep across the OWASP Benchmark v1.2 dataset.}
  \label{fig:overall}
\end{figure}

Table~\ref{tab:overall} presents the aggregate performance metrics for QASecClaw and the Semgrep baseline on the complete OWASP Benchmark v1.2. Fig.~\ref{fig:overall} provides a visual comparison.

\begin{table}[!b]
\centering
\caption{Overall Performance on OWASP Benchmark v1.2 (2,740 test cases)}
\label{tab:overall}
\begin{tabular}{lccccc}
\toprule
\textbf{Tool} & \textbf{Prec.} & \textbf{Recall} & \textbf{F1} & \textbf{FPR} & \textbf{$J$} \\
\midrule
QASecClaw & \textbf{0.951} & 0.871  & \textbf{0.909} & \textbf{0.048} & \textbf{0.823} \\
Semgrep   & 0.695          & \textbf{0.900} & 0.784          & 0.423          & 0.477 \\
\bottomrule
\end{tabular}
\end{table}

QASecClaw achieves an F1-score of \textbf{0.909}, representing a \textbf{16.0\%} improvement over Semgrep's 0.784. The improvement is predominantly driven by a dramatic increase in Precision from 0.695 to 0.951, while Recall experiences a modest decrease from 0.900 to 0.871 (a 3.2\% relative reduction). The False Positive Rate drops from 0.423 to 0.048 — a reduction of \textbf{88.6\%}. Correspondingly, Youden's J increases from 0.477 to 0.823, indicating substantially better discriminative ability.

\noindent \textbf{Observation for RQ1:} QASecClaw outperforms standalone Semgrep on the OWASP Benchmark, achieving a 16.0\% higher F1-score through dramatic false positive reduction with minimal recall loss.

\subsection{RQ2: False Positive Reduction Analysis}

\begin{table}[!t]
\centering
\caption{Confusion Matrix Comparison}
\label{tab:confusion}
\begin{tabular}{lrrrr}
\toprule
\textbf{Tool} & \textbf{TP} & \textbf{FP} & \textbf{TN} & \textbf{FN} \\
\midrule
QASecClaw & 1,233 & \textbf{64}  & \textbf{1,261} & 182 \\
Semgrep   & 1,273 & 560          & 765            & 142 \\
\midrule
\textbf{$\Delta$} & $-$40 & $-$496 & +496 & +40 \\
\bottomrule
\end{tabular}
\end{table}

Table~\ref{tab:confusion} presents the confusion matrix comparison. QASecClaw correctly reclassified \textbf{496 false positives} as true negatives, while only misclassifying 40 true positives as false negatives. This yields a false positive suppression ratio of 88.6\% ($496/560$) with a true positive loss ratio of only 3.1\% ($40/1273$).

\begin{figure}[!t]
  \centering
  \includegraphics[width=\columnwidth]{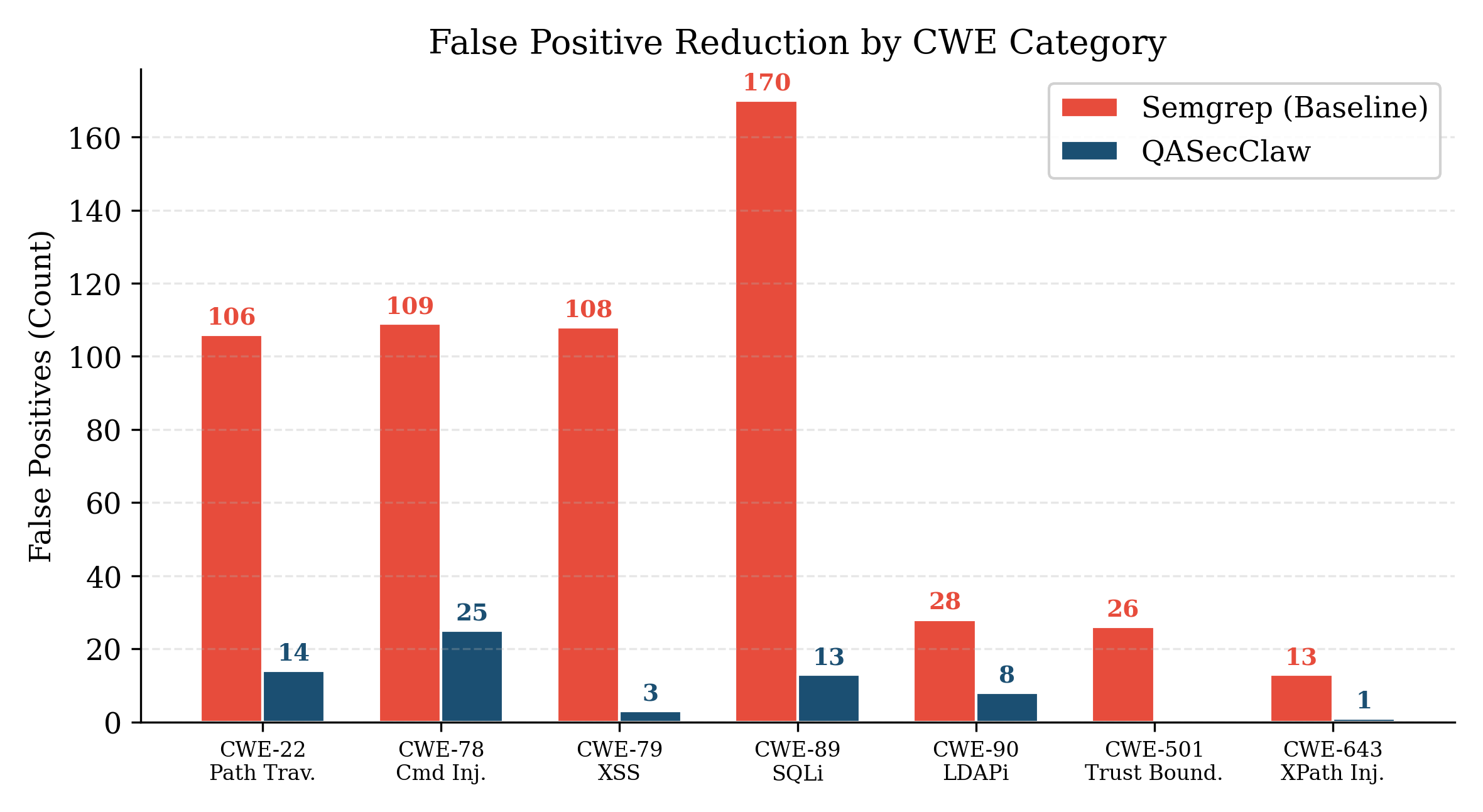}
  \caption{Per-CWE false positive counts for Semgrep vs.\ QASecClaw. QASecClaw reduces false positives across all affected CWE categories.}
  \label{fig:fp_reduction}
\end{figure}

Fig.~\ref{fig:fp_reduction} illustrates the per-CWE false positive reduction, showing consistent improvements across all vulnerability categories that exhibited false positives in the baseline.

\noindent \textbf{Observation for RQ2:} QASecClaw's LLM-augmented filtering eliminates 88.6\% of false positives produced by standalone SAST, with a favorable trade-off of only 3.1\% recall loss.

\subsection{RQ3: Per-CWE Performance Analysis}

\begin{figure*}[!t]
  \centering
  \includegraphics[width=0.85\textwidth]{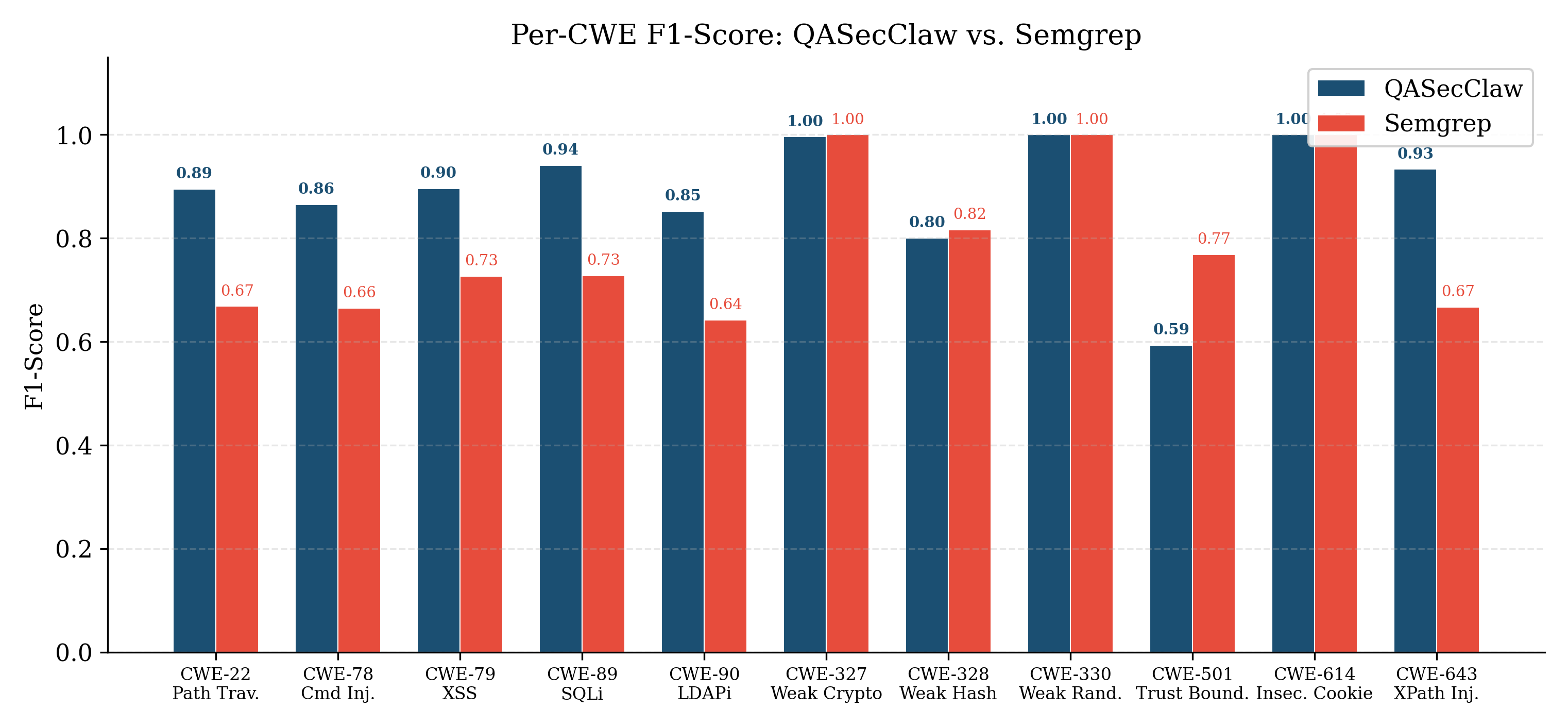}
  \caption{Per-CWE F\textsubscript{1}-score comparison. QASecClaw achieves substantial F\textsubscript{1} improvements on injection-class CWEs while maintaining parity on cryptographic CWEs.}
  \label{fig:per_cwe_f1}
\end{figure*}

\begin{figure}[!t]
  \centering
  \includegraphics[width=\columnwidth]{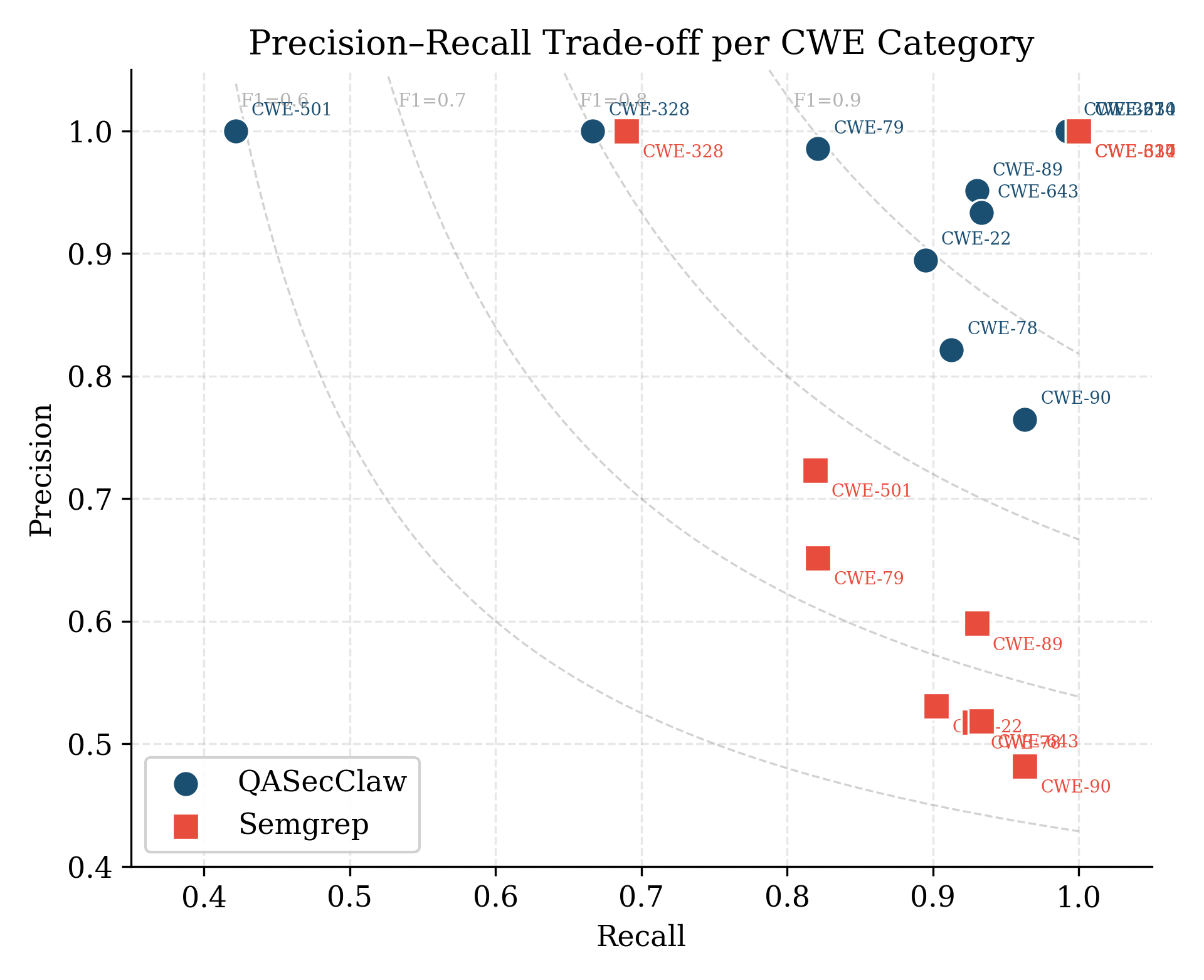}
  \caption{Precision--Recall scatter plot per CWE category. QASecClaw points cluster in the high-precision, high-recall region (upper right), while Semgrep points exhibit high recall but low precision for injection CWEs.}
  \label{fig:pr_scatter}
\end{figure}

Table~\ref{tab:per_cwe} presents the detailed per-CWE breakdown. Fig.~\ref{fig:per_cwe_f1} and Fig.~\ref{fig:pr_scatter} provide visual comparisons.

\begin{table*}[t]
\centering
\caption{Per-CWE Performance Breakdown: QASecClaw vs.\ Semgrep}
\label{tab:per_cwe}
\begin{tabular}{ll|ccc|ccc|cc}
\toprule
& & \multicolumn{3}{c|}{\textbf{QASecClaw}} & \multicolumn{3}{c|}{\textbf{Semgrep}} & \multicolumn{2}{c}{\textbf{$\Delta$ F1}} \\
\textbf{CWE} & \textbf{Category} & \textbf{Prec.} & \textbf{Rec.} & \textbf{F1} & \textbf{Prec.} & \textbf{Rec.} & \textbf{F1} & \textbf{Abs.} & \textbf{Rel.} \\
\midrule
CWE-22  & Path Traversal       & 0.895 & 0.895 & 0.895 & 0.531 & 0.902 & 0.669 & +0.226 & +33.8\% \\
CWE-78  & Command Injection    & 0.821 & 0.913 & 0.865 & 0.518 & 0.929 & 0.665 & +0.200 & +30.1\% \\
CWE-79  & XSS                  & 0.985 & 0.821 & 0.896 & 0.652 & 0.821 & 0.727 & +0.169 & +23.3\% \\
CWE-89  & SQL Injection        & 0.951 & 0.930 & 0.941 & 0.598 & 0.930 & 0.728 & +0.213 & +29.2\% \\
CWE-90  & LDAP Injection       & 0.765 & 0.963 & 0.853 & 0.482 & 0.963 & 0.642 & +0.211 & +32.9\% \\
CWE-327 & Weak Cryptography    & 1.000 & 0.992 & 0.996 & 1.000 & 1.000 & 1.000 & $-$0.004 & $-$0.4\% \\
CWE-328 & Weak Hashing         & 1.000 & 0.667 & 0.800 & 1.000 & 0.690 & 0.817 & $-$0.017 & $-$2.0\% \\
CWE-330 & Weak Randomness      & 1.000 & 1.000 & 1.000 & 1.000 & 1.000 & 1.000 & 0.000 & 0.0\% \\
CWE-501 & Trust Bound.\ Viol.  & 1.000 & 0.422 & 0.593 & 0.723 & 0.819 & 0.768 & $-$0.175 & $-$22.8\% \\
CWE-614 & Insecure Cookie      & 1.000 & 1.000 & 1.000 & 1.000 & 1.000 & 1.000 & 0.000 & 0.0\% \\
CWE-643 & XPath Injection      & 0.933 & 0.933 & 0.933 & 0.519 & 0.933 & 0.667 & +0.266 & +39.9\% \\
\bottomrule
\end{tabular}
\end{table*}

\textbf{Injection vulnerabilities} (CWE-22, 78, 79, 89, 90, 643) show the largest improvements, with F1-score gains ranging from +0.169 (XSS) to +0.266 (XPath Injection). These CWEs are historically problematic for SAST tools because the distinction between exploitable and safe code depends on subtle data flow analysis that pattern-matching rules struggle to capture.

\textbf{Cryptographic vulnerabilities} (CWE-327, 328, 330) maintain near-parity with the baseline. This is expected because Semgrep's rules for these categories already achieve near-perfect precision (both tools report 0 false positives), meaning there are no false positives for the LLM filter to correct.

\textbf{CWE-501 (Trust Boundary Violation)} is the only category where QASecClaw's F1-score is notably lower than the baseline ($-$22.8\%). Investigation reveals that the LLM filter was overly conservative for this category, correctly eliminating all 26 false positives from Semgrep but also misclassifying 48 true positives as benign. This suggests that trust boundary violations require more specialized prompting strategies.

\noindent \textbf{Observation for RQ3:} QASecClaw delivers the largest improvements on injection-class CWEs (F1 gains of +17\% to +40\%), maintains parity on well-detected cryptographic CWEs, and shows room for improvement on trust boundary violations.

\section{Discussion and Limitations}
\label{sec:discussion}

\subsection{Practical Implications}
The main practical value of QASecClaw is its ability to reduce the manual triage burden created by noisy Static Application Security Testing (SAST) output. In the OWASP Benchmark evaluation, the number of false positives decreases from 560 to 64, corresponding to an 88.6\% reduction. In a real Continuous Integration and Continuous Deployment (CI/CD) pipeline, this reduction can save substantial developer and security analyst time because fewer warnings need manual inspection.

Beyond time savings, the more important implication is developer trust. When a tool repeatedly reports false alarms, developers become less likely to inspect its output carefully. This creates the risk that real vulnerabilities will be ignored because they are mixed with many incorrect warnings. By increasing precision from 0.695 to 0.951, QASecClaw makes each reported finding more likely to be actionable. This can encourage developers to treat SAST output as a useful security signal rather than background noise.

\subsection{Why LLM-Based Filtering Helps}
The results suggest that QASecClaw works because it separates vulnerability discovery from vulnerability verification. Semgrep remains responsible for broad candidate generation, which preserves the high-recall behavior expected from SAST tools. The coding-specialized Large Language Model (LLM) is then used in a narrower role: it reviews already reported findings with source-code context and decides whether the warning is likely to be real.

This division of labor is important. A rule-based SAST engine can identify suspicious syntactic patterns, but it often cannot fully interpret whether surrounding code makes the pattern safe. For example, a finding may depend on whether user input is sanitized, whether a query is parameterized, whether a path is canonicalized, or whether output is encoded before rendering. These are context-sensitive judgments. The SAST Filter Agent improves precision by checking this surrounding context before retaining a finding. Therefore, the improvement is not because the LLM replaces SAST, but because the LLM performs a constrained verification task after SAST has already narrowed the search space.

\subsection{Failure Cases and CWE-501}
The results also show that LLM-based filtering is not uniformly effective across all vulnerability categories. QASecClaw performs strongly on injection-related categories, where the key question is often whether untrusted input reaches a dangerous sink without proper sanitization. However, it performs worse on CWE-501 Trust Boundary Violation. In this category, QASecClaw removes false positives but also suppresses several true positives, which lowers recall.

This behavior suggests that trust-boundary violations are harder for the current prompt design. Unlike injection vulnerabilities, where the source, transformation, and sink are often visible in code, trust-boundary violations can depend on application design, framework assumptions, session handling, and security policy. The LLM may apply a stricter interpretation than the benchmark label expects, treating some cases as benign when they are considered vulnerable by the benchmark. This indicates that future versions of the SAST Filter Agent should use CWE-specific prompting, examples, or fallback rules for categories where the definition is more policy-dependent.

\subsection{Cost--Accuracy Trade-off}
The improvement comes with additional computational and operational cost. QASecClaw uses a coding-specialized LLM to review SAST findings, and this introduces inference latency and API cost. In our evaluation, approximately 1,833 unique test configurations were processed by the SAST Filter Agent in batches of 15, requiring approximately 122 LLM calls. The full pipeline, including Semgrep scanning, LLM-based filtering, and scoring, completed in approximately 45 minutes on a standard development workstation.

For practical deployment, this cost should be compared with the human cost of manual triage. In many settings, a small LLM inference cost may be justified if it removes a large number of false positives before developers review the report. However, for very large codebases, additional optimizations may be necessary. These include caching LLM decisions for unchanged files, filtering only medium-confidence or high-risk findings, and using different batch sizes based on file length and context-window constraints.

\subsection{Fail-Open Safety Design}
A key design decision in QASecClaw is the fail-open policy. If the LLM fails to return a valid response because of an API error, rate limit, timeout, malformed output, or network interruption, QASecClaw retains all findings in the affected batch. This design prevents transient LLM failures from silently suppressing real vulnerabilities.

This conservative behavior is important because false positive reduction should not come at the cost of unsafe vulnerability removal. In our evaluation, approximately 10 to 15 batches experienced failures, mainly because of network interruptions. The fail-open mechanism retained those findings, which may have slightly increased the final false positive count. Therefore, the reported performance should be interpreted as conservative rather than overly optimistic.

\subsection{Limitations}
This study has several limitations. First, the evaluation is based on the OWASP Benchmark v1.2, which provides controlled and reproducible ground truth but consists of synthetic Java test cases. These cases are useful for comparing security tools, but they may not fully represent the complexity of real production systems, where vulnerabilities can involve larger codebases, framework behavior, configuration files, and multi-file dependencies.

Second, the current evaluation is limited to Java web application test cases. The effectiveness of QASecClaw on other programming languages, frameworks, and application domains remains to be evaluated. Since SAST rules, vulnerability patterns, and language-specific sanitization mechanisms vary across ecosystems, generalization should not be assumed without further experiments.

Third, the baseline comparison uses Semgrep only. Semgrep is a suitable open-source baseline because it supports Java and produces structured output, but additional comparisons with tools such as SonarQube, Fortify, Checkmarx, and SpotBugs would provide a broader view of the approach's effectiveness.

Fourth, LLM outputs may vary across model versions, prompts, and deployment environments. Although we reduce this risk by using deterministic settings, structured output parsing, and fail-open handling, future work should include repeated runs to measure variance and compare multiple coding-specialized LLMs.

Finally, the scoring follows the OWASP Benchmark file-level matching convention using the \texttt{BenchmarkTestNNNNN} identifier. This is consistent with the benchmark methodology, but it does not always distinguish whether a tool reported the exact intended vulnerability instance inside the file. Instance-level matching would provide a more fine-grained evaluation but would require additional annotation and tool-output normalization.

\section{Conclusion and Future Work}
\label{sec:conclusion}

This paper presented QASecClaw, a multi-agent approach that improves Static Application Security Testing (SAST) by adding coding-specialized Large Language Model (LLM)-based contextual code review to the security analysis pipeline. The central goal of QASecClaw is not to replace conventional SAST tools, but to make their output more useful by reducing false positives that waste developer effort and weaken trust in automated security scanners. In the proposed approach, Semgrep first produces candidate findings, and the SAST Filter Agent then reviews each finding with source-code context to decide whether it should be retained or suppressed. The evaluation on the complete Open Worldwide Application Security Project (OWASP) Benchmark v1.2, containing 2,740 Java test cases across 11 Common Weakness Enumeration (CWE) categories, shows that QASecClaw improves the F\textsubscript{1}-score from 0.784 to 0.909, reduces false positives by 88.6\%, and introduces only a 3.1\% reduction in recall. These results suggest that LLM-augmented verification can be a practical way to improve the signal quality of SAST tools while keeping the conservative behavior needed for security analysis. Future work will extend the evaluation to real-world open-source projects with known Common Vulnerabilities and Exposures (CVEs), compare QASecClaw with additional SAST engines and dynamic analysis tools, refine prompting strategies for under-performing categories such as CWE-501, evaluate multiple coding-specialized LLMs, and study the cost-effectiveness of the approach in enterprise-scale development environments.

\section*{Acknowledgments}
The authors thank the OWASP Foundation for making the Benchmark Project publicly available and the Semgrep maintainers for their open-source SAST engine. We also acknowledge the developers of OpenClaw for providing the agent runtime upon which QASecClaw is built. This work was supported in part by the Department of Computer Science and Engineering, University of Rajshahi.

\bibliographystyle{IEEEtran}
\bibliography{qasecclaw_refs}
\end{document}